\documentclass[pdflatex,sn-nature]{sn-jnl}

\usepackage{graphicx}%
\usepackage{amsmath,amssymb,amsfonts}%
\usepackage{xcolor}%
\usepackage{textcomp}%
\usepackage{booktabs}%

\unnumbered

\begin{document}

\title{Interfacial Water Responds Linearly to Charge yet Is Charge-Asymmetric}

\author*[1,2]{\fnm{Yair} \sur{Litman}}\email{litmany@mpip-mainz.mpg.de}

\author*[1,2]{\fnm{Yongkang} \sur{Wang}}\email{wangy3@mpip-mainz.mpg.de}

\author[3]{\fnm{Stephen J.} \sur{Cox}}

\author*[1,2]{\fnm{Mischa} \sur{Bonn}}\email{bonn@mpip-mainz.mpg.de}

\affil[1]{\orgname{Max Planck Institute for Polymer Research}, \orgaddress{\street{Ackermannweg 10}, \city{Mainz}, \postcode{55128}, \country{Germany}}}

\affil[2]{\orgname{Max Planck Initiative for Liquid Research}, \orgaddress{\street{Ackermannweg 10}, \city{Mainz}, \postcode{55128}, \country{Germany}}}

\affil[3]{\orgdiv{Department of Chemistry}, \orgname{Durham University}, \orgaddress{\street{South Road}, \city{Durham}, \postcode{DH1 3LE}, \country{United Kingdom}}}

\abstract{
The organization of interfacial water shapes the electric double layer behavior in aqueous electrolytes. Asymmetries upon charge reversal are widely observed and are commonly attributed to specific ion adsorption or surface chemistry. Yet it is unresolved whether interfacial water is net oriented at zero charge, and whether such orientation alone makes its response charge-asymmetric. Here, using electrochemical heterodyne-detected sum-frequency generation spectroscopy and constant-potential molecular dynamics simulations, we resolve the absolute orientation of water at a weakly interacting graphene electrode. We show that interfacial water retains a net H-down orientation, with hydrogens pointing toward the bulk, at zero net surface charge. Superimposed on this offset, the field-induced reorientation is linear and symmetric upon charge reversal, independent of how the charge is introduced or distributed. The zero-charge offset therefore shifts the potential of zero water orientation negatively relative to the potential of zero charge, making the interfacial water response intrinsically charge-asymmetric. These results establish an electrostatic baseline against which extrinsic contributions at electrified interfaces can be isolated.

}

\keywords{Interfacial water, Electrified interfaces, Sum-frequency generation spectroscopy, Graphene}

\maketitle

\section{Main}

The unequal response of a system to positive and negative charges of equal magnitude, usually referred to as charge asymmetry, is ubiquitous in aqueous interfacial chemistry and biology. Its manifestations are evident even at macroscopic interfaces: in electrowetting, water on hydrophobic dielectrics responds differently to positive and negative voltages~\cite{verheijen_1999_reversible}, while supercooled water freezes at higher temperatures on positively than on negatively charged pyroelectric surfaces~\cite{ehre_2010_water}. Beyond planar interfaces, like-charged colloids interact in a sign-dependent manner: in water, negatively charged particles attract one another at long range, whereas positively charged ones repel~\cite{wang_2024_charge}. In biology, protein stability and solubility likewise depend asymmetrically on the sign of dissolved ions, as reflected in the distinct anionic and cationic Hofmeister series~\cite{okur_2017_beyond, jungwirth_2014_beyond}. Specific mechanisms have been identified in some of these cases, including trapped charge in electrowetting~\cite{verheijen_1999_reversible, jones_2005_electromechanical}, ion partitioning in electrofreezing~\cite{javitt_2022_chemical}, and distinct binding sites for anions and cations on proteins~\cite{okur_2017_beyond}. Yet it remains unresolved whether, and to what extent, water itself accounts for this charge asymmetry beyond extrinsic contributions such as specific ion adsorption and surface chemistry.

The net orientation of interfacial water dipoles contributes directly to the potential drop across the electric double layer and to the local field~\cite{Shin_NatComm_2022, Huang_JPCC_2021, garciaaraez_2009_potential, wang_2025_beyond}. 
If water behaved as a linear dielectric continuum, unpolarized at zero charge, this orientation would respond identically to either sign of charge. This picture, however, breaks down at interfaces~\cite{Gonella_ChemRev_2021, loche_2018_breakdown, dinpajooh_2016_dielectric, bonthuis_2012_profile}. At metal electrodes, for example, the inner-layer capacitance varies nonlinearly with potential and is asymmetric about the potential of zero charge (PZC)~\cite{Grahame_1947_EDL}, and at Pt(111) both the nonlinearity and the asymmetry have been traced to potential-dependent chemisorption of water~\cite{Le_sciadv_2020}.
More generally, charge asymmetry at aqueous interfaces has often been ascribed, at least in part, to water itself~\cite{scheu_2014_charge, Montenegro_Nat_2021, urashima_2018_topmost, Wei_JACS_2023, walker_2022_role, bin_2026_resist, sundararaman_2023_solvent, shandilya_2022_interfacial}.

Indeed, the charge distribution of the water molecule is asymmetric, leading to distinct solvation of anions and cations: anions orient O-H bonds, making water point one of its hydrogen atoms toward anions, but cations align the static dipole: water points its oxygen toward cations~\cite{hummer_1996_free, mobley_2008_charge, tielrooij_2011_anisotropic, Cox_JCP_2021}.
The response of water to charge could therefore be intrinsically nonlinear, with a different magnitude for each sign. At charged surfaces and electrodes, however, this response is entangled with hydrogen bonding to surface groups~\cite{urashima_2018_topmost, Wei_JACS_2023, speelman_2025_quantifying, rehl_2022_water}, chemisorption~\cite{Mei_JCPL_2021, Mei_JCP_2024, Le_sciadv_2020}, surface-specific ordering of the water H-bond network~\cite {bin_2026_resist, sundararaman_2023_solvent}, or ion-specific effects~\cite{Gonella_ChemRev_2021, Litman_NatChem_2024, javitt_2022_chemical}. Because ion distributions, surface chemistry, and water structure all change together upon charging, the response of water itself is hard to isolate.
This leaves a fundamental question open: is the charge asymmetry of water at extended interfaces a result of the asymmetric molecular water response to opposite charges? In other words: does the charge asymmetry of interfacial water require a nonlinear response to charge, or does it arise from a linear response superimposed on an orientation already present at zero charge?

Answering this question requires two ingredients: an interface at which the surface charge can be tuned while extrinsic contributions are minimized, and a probe that resolves the absolute orientation of interfacial water under varied surface charge~\cite{li_2021_gate, wang_2019_gate, liu_2014_situ, wallentine_2020_plasmon} together with the net charge it experiences. Graphene offers such a limiting case~\cite{Montenegro_Nat_2021, Wang_Nature_2023, Wang_Angewandte_2023, Xu_Nature_2023, wang_spectral_2025, yang_2022_nature, achtyl_2015_aqueous, achtyl_2013_free}. It is chemically inert and does not chemisorb water, and simple ions such as Na$^+$ and Cl$^-$, although they accumulate at its surface, barely perturb the interfacial water structure~\cite{mccaffrey_2017_mechanism, advincula_JACS_2026_breaking, wang_2025_beyond, jiang_2016_molecular}. The response of interfacial water is therefore free of chemisorption and only weakly affected by the electrolyte. Probing this response directly, however, requires access to the absolute orientation of water at a buried electrode together with an accurate determination of the interfacial charge~\cite{Montenegro_Nat_2021, Wang_Nature_2023, Wang_Angewandte_2023, Xu_Nature_2023, wang_spectral_2025}. Heterodyne-detected vibrational sum-frequency generation (HD-VSFG) spectroscopy provides the former, as the sign of Im$\chi^{(2)}$ encodes the net orientation of interfacial water~\cite{Nihonyanagi_flip_flop_2009}, and, through its dependence on electrolyte concentration, the latter~\cite{wen_unveiling_2016, Seki_Chem_2021, Wang_Nature_2023}. 

Here, using electrochemical HD-VSFG spectroscopy on a CaF$_2$-supported graphene electrode combined with molecular dynamics (MD) simulations, we show that interfacial water retains a net H-down orientation toward bulk water away from the electrode at the potential of zero net surface charge. Superimposed on this offset, the field-induced reorientation varies linearly with net charge and reverses symmetrically upon charge reversal, irrespective of whether the charge is varied electrochemically or through pH, and of whether it resides on the substrate or on graphene. As a consequence, net interfacial water alignment becomes unoriented only at a finite negative charge, displacing the potential of zero water orientation from the potential of zero net charge and rendering the total interfacial water response charge-asymmetric, without invoking specific ion adsorption or water chemisorption at the graphene surface. MD simulations reproduce this linear yet charge-asymmetric response across a range of water--graphene interaction strengths and for several water models. These results establish the electrostatic baseline of interfacial water, against which extrinsic contributions at more complex electrodes can be identified.

\section{Results and Discussion}

\subsection{Interfacial Water Reorganization at Electrified Interfaces}

\begin{figure}\centering
\includegraphics[width=.85\textwidth]{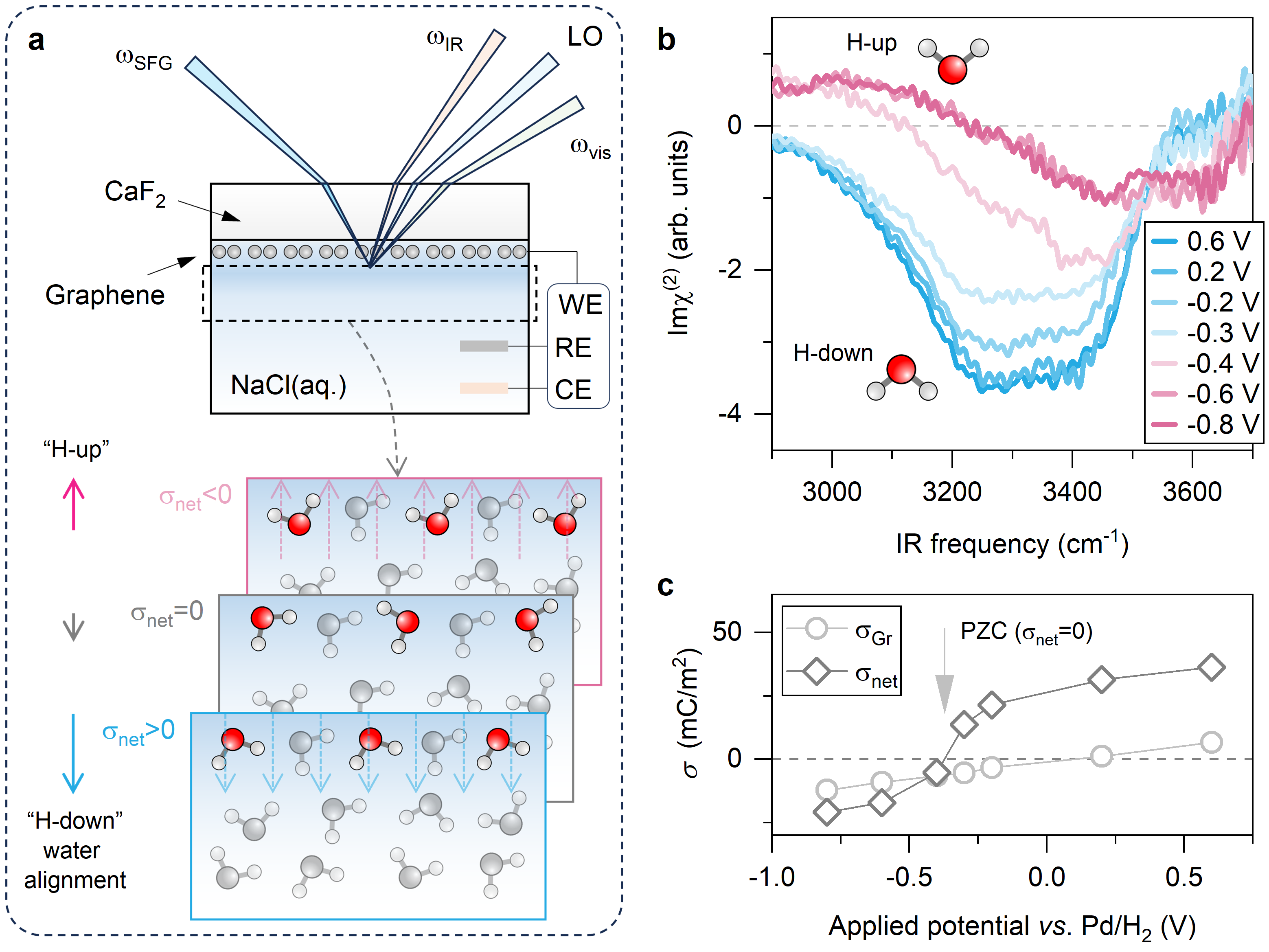}
  \caption{\textbf{Charge-induced reorientation of interfacial water at an electrified graphene/water interface under electrochemical conditions revealed by HD-VSFG spectroscopy.} 
  \textbf{a}. Schematic representation of the CaF$_2$-supported graphene/water interface in 2 M NaCl solution, with ions omitted for clarity. The HD-VSFG beam geometry, including the generated SFG signal ($\omega_{\mathrm{SFG}}$), local oscillator (LO), visible ($\omega_{\mathrm{vis}}$), and infrared ($\omega_{\mathrm{IR}}$) beams, is indicated. The lower panels show molecular diagrams of interfacial water under opposite net surface charges, with dashed arrows indicating the corresponding electric fields and illustrating that a symmetric field-induced reorientation, added to a zero-charge offset, yields an asymmetric absolute orientation.
  \textbf{b}. $\mathrm{Im}\chi^{(2)}$ spectra of the interfacial water O--H stretching region at CaF$_2$-supported graphene/water interfaces under applied electrochemical potentials. The open-circuit potential (OCP) of graphene is approximately 0.2~V versus Pd/H$_2$. Measurements were performed using a three-electrode configuration, with graphene, Au, and Pd/H$_2$ serving as the working, counter, and reference electrodes, respectively. The insets show molecular diagrams of H-up and H-down interfacial water near the graphene electrode.
  \textbf{c}. Potential-dependent net surface charge at the CaF$_2$-supported graphene/water interface, extracted from comparison of HD-VSFG spectra at different NaCl concentrations using a Gouy--Chapman--Stern (GCS) mean-field approach. The graphene charge density ($\sigma_{\mathrm{Gr}}$), independently determined from Raman G-band spectra, is shown for comparison.
  }
  \label{fig:SFG}
\end{figure}

HD-VSFG provides direct access to the complex second-order nonlinear susceptibility $\chi^{(2)}$ of interfacial water, where the imaginary component, Im$\chi^{(2)}$, encodes the absolute orientation and hydrogen-bonding environment of interfacial water molecules~~\cite{Nihonyanagi_flip_flop_2009}. Access to the complex $\chi^{(2)}$ response enables additive separation of spectral contributions~~\cite{Ji_PRL_2008}, facilitating quantitative analysis of electrode charging through appropriate modeling of the charged interface~~\cite{Wang_Nature_2023, Seki_Chem_2021, fellows_2024_sum}. 
In this work, we use high ionic strength (2 M NaCl) to suppress long-range electrostatic contributions, so the SFG response predominantly reflects the structure of interfacial water within the top one to two molecular layers~~\cite{Wei_JACS_2023, wen_unveiling_2016, fellows_2024_sum}. We used NaCl because it weakly adsorbs on graphene~~\cite{mccaffrey_2017_mechanism, advincula_JACS_2026_breaking, jiang_2016_molecular}, minimizing direct ion-induced perturbations of interfacial water. Figure \ref{fig:SFG}a shows a schematic of the HD-VSFG experimental setup. Further details on sample preparation and HD-VSFG measurements are provided in Refs.~~\cite{Montenegro_Nat_2021, Wang_Angewandte_2023, Wang_Nature_2023, Dreier_JPCC_2019} and in the Methods section.

Figure \ref{fig:SFG}b presents the measured $\mathrm{Im}\chi^{(2)}$ spectra as a function of applied potential ($-0.8$ to $+0.6$ V vs. Pd/H$_2$) at the graphene electrode using a standard three-electrode electrochemical configuration, with the bulk solution initially adjusted to pH 6. At positive potentials, the spectrum exhibits a broad negative band spanning 2900--3550 cm$^{-1}$, reflecting the distribution of hydrogen-bonding configurations sampled by interfacial water molecules. The negative sign indicates a net H-down orientation of hydrogen-bonded O--H groups, with their hydrogen atoms pointing toward the bulk solution. Upon decreasing the applied potential, the amplitude of this hydrogen-bonded O--H band gradually decreases, indicating a reduction in the net H-down orientational alignment of interfacial water.

In addition to the reorganization of interfacial water at the exposed graphene surface, a negative feature at 3630 cm$^{-1}$~emerges at potentials below $-0.3$ V. While previous conventional SFG studies have assigned this feature to dangling O--H groups of interfacial water~~\cite{Montenegro_Nat_2021}, our HD-VSFG spectra indicate that this negative response is inconsistent with that assignment, as this would require its sign to be positive. This response is consistent with the O--H stretching vibration of hydroxylated calcium fluoride (Ca--O--H). These Ca--O--H groups originate from adsorbed hydroxide species~~\cite{Becraft_Langmuir_2001}, which can form at the CaF$_2$/graphene interface due to a small population of water molecules trapped between the substrate and graphene~~\cite{Montenegro_Nat_2021, Wang_Angewandte_2023, Wang_Nature_2023, wang_NatComm_2025}. The emergence of this feature at negative potentials is consistent with potential-induced water dissociation at the graphene/water interface, which increases the local pH, followed by proton transfer through the graphene layer that promotes hydroxylation of the underlying CaF$_2$ surface~\cite{Wang_Angewandte_2023}. Notably, water confined between graphene and the substrate can also contribute to the SFG response; however, this contribution remains minor compared with that from water at the exposed graphene/water interface (see Supplementary Information (SI), Figure S2). Therefore, the potential-dependent spectral evolution in the H-bonded O--H stretching region can be attributed predominantly to the reorganization of interfacial water at the exposed graphene surface.

\subsection{Charge-Induced Interfacial Water Reorientation Response is Linear}

Qualitatively, the change in interfacial-water alignment, as reflected by the spectral amplitude of the H-bonded O–H peak, exhibits a strongly nonlinear dependence on the applied potential, despite the approximately linear charging response of the graphene electrodes, independently established from well-established Raman G-band frequency analysis (Figs. \ref{fig:SFG}b and \ref{fig:SFG}c, see also Methods)~~\cite{das_2008_monitoring, Montenegro_Nat_2021, Wang_FaradayDiscuss_2024, Wang_Nature_2023}. This apparent nonlinearity arises because the interfacial electric field at the exposed graphene surface is not determined solely by electrochemical charging of graphene but also by the surface charges of the supporting CaF$_2$ substrate, which reorients interfacial water through charge-induced polarization of graphene (related to the so-called 'wetting transparency' of graphene)~~\cite{KIM_Chem_2021, wang_2026_wetting, hou_2026_machine,Rafiee_NatMat_2012}. CaF$_2$ is positively charged at neutral pH, with an isoelectric point in the range of pH 9–10~~\cite{khatib_water_2016,Becraft_Langmuir_2001}. Importantly, potential-dependent hydroxylation of the CaF$_2$ surface progressively neutralizes this surface charge. Consequently, under applied potentials in electrochemical HD-VSFG measurements, the interfacial electric field reflects contributions from both capacitive charging of graphene and potential-dependent pseudocapacitive surface charging of the CaF$_2$ substrate~~\cite{KIM_Chem_2021, Wang_Angewandte_2023}. In fact, the net surface charge density experienced by interfacial water, extracted from HD-VSFG spectra using a mean-field Gouy–Chapman–Stern (GCS) framework following established protocols~~\cite{wen_unveiling_2016, Wang_Nature_2023, Seki_Chem_2021}, exhibits a nonlinear variation from approximately +36 to -21 mC/m$^{2}$, with the potential of zero (net) charge at $\sim$-0.3 to -0.4 V (Figure \ref{fig:SFG}c; see also Figure S1 for additional experimental data and the Methods for details of the data analysis), in agreement with previous studies~~\cite{Wang_Nature_2023, Wang_Angewandte_2023, wang_2026_wetting}.

Quantitatively, the interfacial water response, obtained by integrating the $\mathrm{Im}\chi^{(2)}$ spectra over 2900–3550 cm$^{-1}$, scales approximately linearly with the net surface charge across the entire investigated range, within experimental uncertainty (Figure \ref{fig:orientations}a). We focus on this spectral region because it captures the overall orientational response of hydrogen-bonded interfacial water. Moreover, moderate variations in the integration range, which account for contributions from interfacial water molecules with different hydrogen-bonding environments, do not qualitatively affect the observed trend (Figure~S3). The uncertainty primarily arises from the maximum possible contribution of nanoconfined two-layer water trapped between the graphene and the substrate~~\cite{wang_NatComm_2025} (Figure~S2).

Although the detailed molecular structure of water within the electric double layer may evolve at high ionic strengths due to shortened Debye lengths, previous studies have shown that the spectral characteristics of field-induced water reorientation remain largely conserved~~\cite{wen_unveiling_2016, fellows_2024_sum, ohno_2017_second, reddy_2018_bulk}. Our observations are consistent with this picture and further explain the robustness of the extracted linear response against the choice of spectral integration window. 

\subsection{Interfacial Water Remains Oriented at the Potential of Zero Net Charge}

The orientation of interfacial water does not vanish at the potential of zero (net) charge. This is evident from the finite SFG response at potentials near -0.3 to -0.4 V (Figs. \ref{fig:SFG}b and \ref{fig:orientations}a), where the inferred net surface charge is approximately zero, yet the spectrum still indicates a preferred H-down orientation of interfacial water. The degree of residual alignment depends on the specific interface, reflecting interfacial symmetry breaking and dispersive interactions to which both graphene and the supporting substrate can contribute~~\cite{shih_2012_breakdown, kong_2018_polarity}. A similar H-down orientation of hydrogen-bonded water has been reported at charge-neutral air/water interfaces~~\cite{Shen_ChemRev_2006, Litman_NatChem_2024}, substrate-free suspended graphene/water interfaces~~\cite{Xu_Nature_2023, wang_2024_heterodyne, wang_spectral_2025, ohto_2018_structure, advincula_JACS_2026_breaking, rashmi_2025_revealing}, and more generally hydrophobic-water interfaces ~\cite{tian_2009_structure, tong_2017_hydrophobic, strazdaite_2015_water, vacha_2011_orientation}. 

The approximately linear dependence of the SFG response on net surface charge further suggests that this residual orientational contribution remains largely unchanged as the surface charge varies. Accordingly, the observed spectral evolution can be described as the superposition of a charge-independent contribution arising from non-electrostatic interfacial interactions and a charge-dependent contribution associated with electrostatically driven water reorientation. The latter varies linearly with the net surface charge and reverses upon charge reversal, indicating that the electrostatic component of interfacial water reorientation is linear and charge-reversal symmetric with respect to surface charge. Consistent with this interpretation, the electrostatically driven response exhibits a similar spectral lineshape, with charge-induced changes primarily reflected in the signal amplitude (see differential spectral analysis in Figure~S5). 

To further validate this relationship, we measured pH-dependent HD-VSFG spectra (Figure S4) and compared them with the potential-dependent measurements, as both perturbations are expected to modulate interfacial water through electrostatic charging, but by distinct mechanisms: the applied potential charges graphene capacitively and the CaF$_2$ substrate pseudocapacitively, whereas varying the pH changes mainly the charge of the CaF$_2$ substrate and leaves the graphene charge largely unchanged.\cite{das_2008_monitoring, Wang_Angewandte_2023, wang_2024_aqueous, Wang_FaradayDiscuss_2024} The pH-dependent measurements likewise exhibit an approximately linear dependence with a nearly identical slope and the same intercept (Figure \ref{fig:orientations}a). These results indicate that, over the range of net surface charge density studied, both the linear response of interfacial water and the charge asymmetry arising from its zero-charge offset are independent of the charging mechanism.

The apparent asymmetric response originally reported for substrate‑supported graphene electrodes by Montenegro et al.\ ~\cite{Montenegro_Nat_2021} can be readily reconciled within this framework: as shown by Wang et al.~\cite{Wang_Nature_2023},  Montenegro et al.\ did not account for potential‑dependent pseudocapacitive charging of the underlying substrate, which is coupled to potential-induced changes in the local pH and consequently modifies the net surface charge. Once this substrate contribution is properly incorporated, the resulting charge-induced water response is consistent with the linear, charge-reversal symmetric behavior established here, indicating that the originally reported asymmetry reflected an incomplete accounting of the interfacial charge rather than a genuine deviation from linear response.

The linear fit in Fig. 2a yields a potential of zero water orientation (PZO) at $\sigma_\mathrm{PZO} = -23 \pm 14~\mathrm{mC/m^{2}}$, corresponding to a shift of more than 0.4 V vs. PZC in applied electrode potential. The uncertainty in $\sigma_\mathrm{PZO}$, $\sim14~\mathrm{mC/m^{2}}$, is inferred from the estimated contribution of confined water. The zero-charge offset is thus comparable in magnitude to the field-induced
reorientation produced by $\sim23~\mathrm{mC/m^{2}}$ of surface charge: at $\sigma = -23~\mathrm{mC/m^{2}}$, interfacial water is essentially unoriented, whereas at
$\sigma = +23~\mathrm{mC/m^{2}}$, its net alignment is roughly twice that at zero charge. Constant-potential simulations place the potential of zero orientation at  $\sim-10~\mathrm{mC/m^{2}}$. The smaller shift likely reflects limitations of the Gouy-Chapman-Stern model, contributions from confined water, and limitations of the employed classical force field, whose fixed point charges neglect the electronic polarization that reduces molecular dipole moments within the first few \AA{} of an aqueous interface~\cite{Jana_JCIM_2024}.

\subsection{Charge Distribution Does Not Shape the Linear Water Response}

Because experimental surface charge densities rely on a Gouy--Chapman--Stern model (see Methods)~~\cite{wen_unveiling_2016, fellows_2024_sum, ohno_2017_second, reddy_2018_bulk, Wang_Nature_2023, Seki_Chem_2021}, we tested the linear but asymmetric response independently using classical molecular dynamics simulations of 1~M NaCl in contact with CaF$_2$-supported graphene, where we directly prescribe the surface charge density. In these simulations, we explore two limiting scenarios in which the surface charge is treated as either entirely localized or entirely delocalized, corresponding to the limiting cases of CaF$_2$ surface charging and graphene charging in our experimental system, respectively. The simulated system consists of a liquid slab confined between two solids, each composed of a monolayer of graphene supported on CaF$_2$. Localized charges were introduced by modifying the charge of selected atoms in the topmost layer of CaF$_2$, while overall charge neutrality was maintained by adding compensating Na$^+$ or Cl$^-$ ions in the aqueous region. 
In all simulations, unless stated otherwise, graphene was kept polarizable using the Siepmann–Sprik constant-potential method~\cite{siepmann_influence_1995}, in which Gaussian charges centered on the carbon atoms are adjusted at each timestep to match a prescribed potential drop between the two electrodes. This potential drop was set to 0.0 V for localized charging and to finite values to induce delocalized charges on graphene. This framework allows independent control of both localized (CaF$_2$) and delocalized (graphene) surface charges.
Additional computational details are provided in the Methods section.

\begin{figure}
\centering
\includegraphics[width=.99\textwidth]{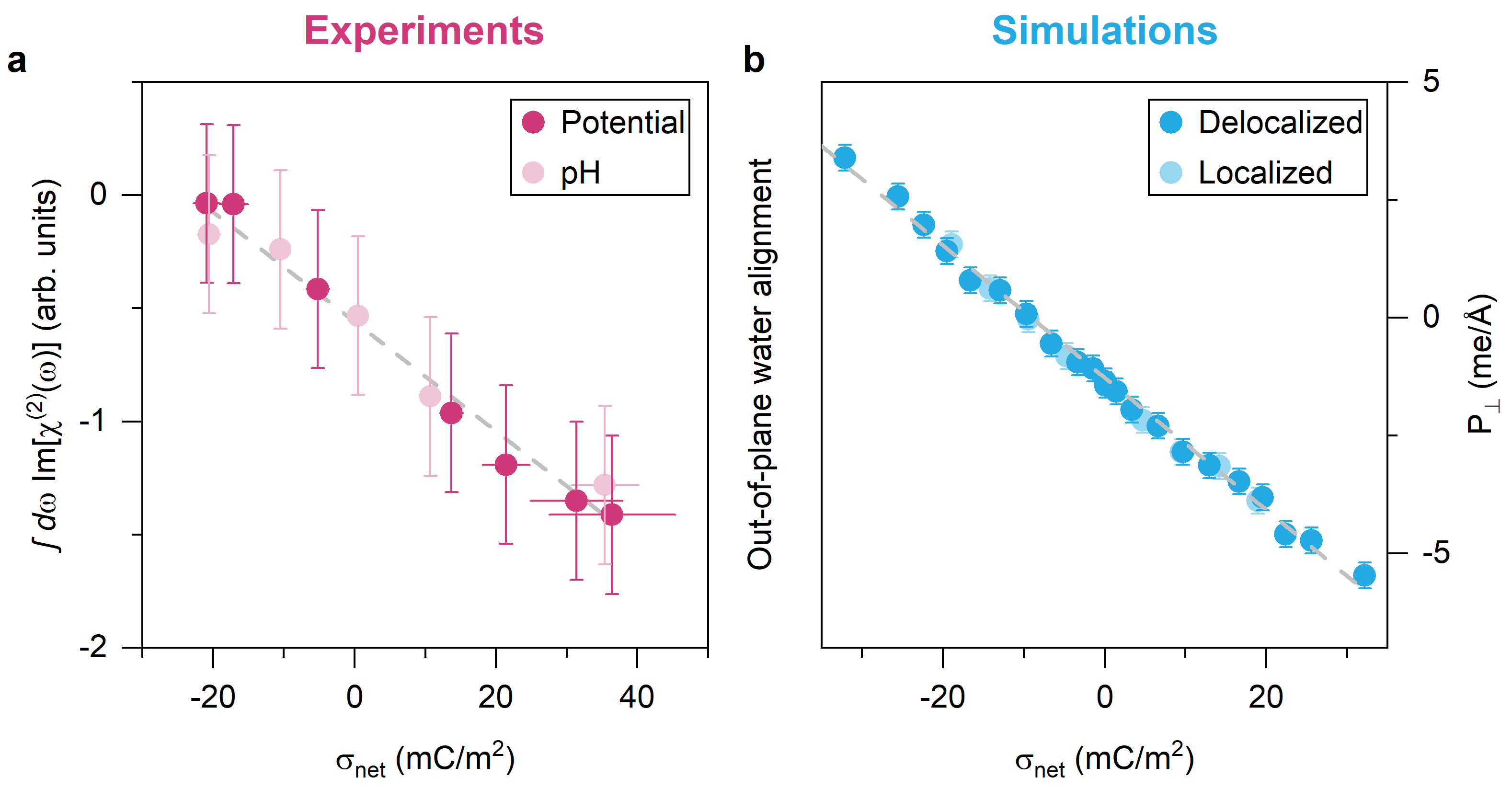}
  \caption{\textbf{Reorientation of interfacial water at an electrified graphene/water interface is linear but charge-asymmetric.} 
  \textbf{a}. Integrated HD-VSFG signal in the hydrogen-bonded O--H stretching region (2900--3550~cm$^{-1}$) as a function of the net surface charge density for measurements under different applied potentials. Corresponding integrated HD-VSFG signals obtained by varying the surface charge through the solution pH are shown for comparison. Horizontal error bars represent the standard deviation of the surface charge density predicted by the Gouy--Chapman--Stern model over the O--H stretching frequency range used for charge extraction. Vertical error bars represent the upper-limit uncertainty associated with the possible contribution from two layers of nanoconfined water (see main text).
  \textbf{b}. Orientation of interfacial water as a function of localized and delocalized surface charge, quantified by the density-weighted orientational order parameter, $P_\perp = -\mu\int \rho(z)\langle  \cos\theta(z) \rangle\,dz$, obtained from molecular dynamics (MD) simulations (see main text).
  Here $\mu$ = 0.489 e\AA~is the SPC/E dipole moment, and the minus sign makes negative $P_\perp$ correspond to net H-down orientation, as for Im$\chi^{(2)}$ in panel a.  
  Grey dashed lines in panels (\textbf{a}) and (\textbf{b}) indicate linear fits to the corresponding data.
  }
  \label{fig:orientations}
\end{figure}

\begin{figure}
\centering
\includegraphics[width=.40\textwidth]{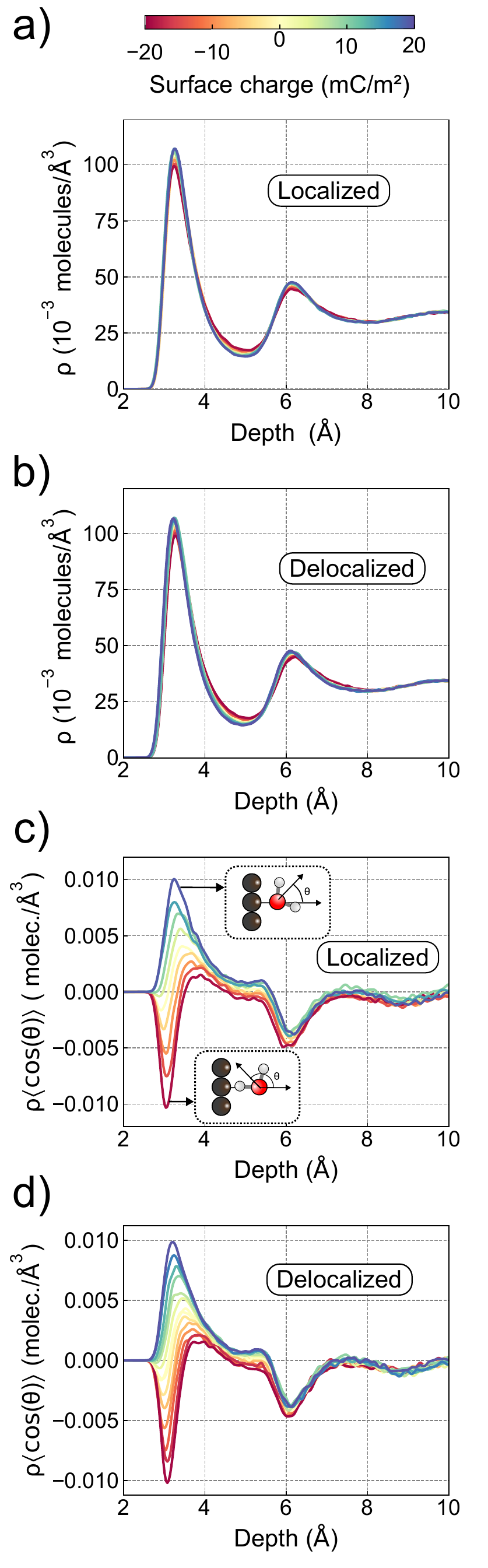}
  \caption{\textbf{Localized and delocalized charges induce similar changes in density and orientational profiles}
  \textbf{a}, \textbf{b}. Oxygen density profiles, $\rho$, at different localized (\textbf{a}) and delocalized (\textbf{b}) surface charge densities, respectively.
  \textbf{c}, \textbf{d}. Orientational profiles, $\langle \cos\theta(z) \rangle$, at different localized (\textbf{c}) and delocalized (\textbf{d}) surface charge densities, respectively. In all cases, the zero depth is defined as the center of the carbon atoms. $\theta$ represents the angle between the direction of the water dipole moment and the surface normal, see insets in panel (\textbf{c}).
  }
  \label{fig:simulation}
\end{figure}

Figure \ref{fig:orientations}b shows the charge-induced reorientation of interfacial water for surface charge densities ranging from $-30$ to $+30$~mC/m$^2$. We quantify the average interfacial water orientation using the density-weighted orientational order parameter, $P_\perp =-\mu \int \rho(z)\langle\cos(\theta(z))\rangle dz$, obtained by integrating the orientation profile across the first two water layers, corresponding to an interfacial region extending approximately 8~\AA{} from the surface. This definition enables direct comparison with the experimental results~~\cite{pop_model_2024},
where $\mu$ is the dipole moment of the water model, $\rho(z)$ represents the oxygen density profile along the surface normal and $\theta(z)$ denotes the angle between the water dipole moment and the surface normal. Figures \ref{fig:simulation}a and \ref{fig:simulation}b present the oxygen density profiles as a function of surface charge for the localized and delocalized charge scenarios, respectively, while the corresponding orientation profiles are shown in Figs. \ref{fig:simulation}c and \ref{fig:simulation}d. Both density and orientation profiles exhibit qualitatively similar trends for localized and delocalized charging. The density profile is only weakly modified over the investigated charge range, with positively charged surfaces inducing a modest enhancement of interfacial layering. In contrast, the water orientation is highly sensitive to small variations in surface charge. The charge-induced orientational ordering is confined within approximately 1~nm of the interface, corresponding to roughly two water layers, while water molecules beyond this region exhibit isotropic, bulk-like behavior. Thus, surface charging primarily modulates the orientation of water molecules within the first two hydration layers, whereas the third hydration layer remains largely unaffected. Positive and negative surface charges induce opposite orientational responses, consistent with the expected electrostatic behavior~~\cite{Olivieri_FarDis_2024}.

The resulting net water orientation shown in Fig. \ref{fig:orientations}b varies linearly and charge-asymmetrically with surface charge for both localized and delocalized charging, in excellent agreement with the experimental observations (Fig. \ref{fig:orientations}a). For delocalized charging, similar behavior has previously been reported at the graphene/water interface at comparable surface charge densities based on the analysis of interfacial water reorientation dynamics~~\cite{Zhan_JPCL_2020}. 
Importantly, the consistency between experiment and simulation strongly suggests that the linear and charge-asymmetric reorientation of interfacial water over the investigated surface charge range is a robust electrostatic response that is largely insensitive to the microscopic distribution of surface charge. Instead, it reflects the collective response of interfacial water to the local electric field. This insensitivity is not unexpected, as graphene acts as a homogenizing layer that suppresses the sensitivity of the average interfacial water structure to the specific spatial distribution~~\cite{wang_2026_wetting}.

\subsection*{Discussion}

Our study focuses on a model system in which extrinsic contributions, such as specific ion adsorption and surface chemistry, are minimized within a relatively low surface charge density regime of tens of mC/m$^{2}$. Graphene provides a chemically well-defined and weakly interacting interface (see  Figures S6 and S7 for further discussion of the ion effect), allowing the intrinsic electrostatic response of interfacial water to be isolated and establishing a reference framework for understanding water organization across diverse material platforms. Under these conditions, our results show that interfacial water responds predominantly through a linear electrostatic mechanism, providing a baseline against which additional chemical contributions can be evaluated. Although we intentionally selected the present system to minimize extrinsic contributions, we anticipate that the underlying electrostatic principles are broadly applicable to other charged aqueous interfaces, provided specific interactions and interfacial heterogeneities are properly considered. In particular, deviations from this intrinsic response are expected in systems where specific ion adsorption, surface chemistry, or molecular specificity play important roles, including technologically and biologically relevant interfaces such as oxide--water, metal--water, and lipid--water interfaces.~\cite{Gonella_ChemRev_2021} Furthermore, the same behavior is recovered for graphene electrodes with different water-carbon van der Waals interaction strengths (Figure~S8), as well as at the electrochemically relevant Au(111) surface for both SPC/E and TIP4P/2005 water (Figure~S9), suggesting that linear electrostatic reorientation represents a general response of interfacial water under weak electrostatic perturbation conditions.
  
A complementary thermodynamic probe of water orientation at the PZC can be obtained by laser-induced temperature-jump experiments that locate the potential of maximum entropy (PME) of double-layer formation~\cite{Climent_JCPB_2002}. On Au(111), the PME lies slightly negative of the PZC, which has been interpreted as a net orientation of interfacial water at zero charge, with the oxygen end directed toward the surface, that is canceled only upon negative charging~\cite{Climent_JCPB_2002}. However, the PME is commonly identified with the potential of zero water orientation, assuming that maximum entropy coincides with vanishing net alignment. Yet the interfacial entropy also includes contributions from the hydrogen-bond network and the ionic double layer.\cite{shandilya_2022_interfacial} 
Indeed, electrochemical microcalorimetry on Au(111) shows that adsorbed ions contribute substantially to the entropy of interface formation, with a clear dependence on the electrolyte ions~\cite {Schonig_PCCP_2023}. HD-VSFG, in contrast, reports the net orientation directly through the sign of Im$\chi^{(2)}$. Combining HD-VSFG with temperature-jump or microcalorimetric measurements on the same electrode would therefore directly test whether the potentials of zero orientation and maximum entropy coincide, and thereby establish whether the PME can serve as a reliable proxy for interfacial water orientation.

Although the charge densities investigated here are relatively low, in the tens of mC/m$^{2}$~range, they overlap with those encountered in many electrochemical, battery, mineral, and biological environments~~\cite{Gonella_ChemRev_2021}. 
The observed linearity is nevertheless expected to break down at high fields, where the dielectric response of water becomes field-dependent and saturates~\cite{kornyshev_1997_nonlocal, Booth_JCP_1951}. Our simulations confirm this above about 60-100 mC/m$^{2}$~(Figures S8 and S9), a regime already challenging to reach experimentally, where water organization becomes asymmetric~\cite{Olivieri_FarDis_2024, Zhang_PCCP_2020, dreier_saturation_2018}. Linearity is also expected to fail where strong non-electrostatic surface–water interactions dominate~\cite{Shin_NatComm_2022, Le_sciadv_2020, walker_2022_role, Mei_JCPL_2021, Mei_JCP_2024}. 
In the SI, we further examine another potential source of nonlinearity arising from multivalent surface charges (Figure S10). We find that replacing monovalent localized surface charges with localized trivalent charges does not alter the linearity of the water response over the range of surface charge densities investigated here.

Our study addresses a fundamental question about the electrostatic response of interfacial water: 
Does its charge asymmetry emerge from an intrinsic nonlinear response, or are the two distinct phenomena? We separated an orientational offset, already present at zero charge, from the linear, field-induced response.
Within the range of surface charge densities studied here, the two are distinct: the offset alone renders interfacial water charge-asymmetric.
This mechanism provides a direct molecular basis for the sign-dependent interactions of like-charged colloids: electrosolvation models attribute the long-range attraction between negatively charged particles in water, and the repulsion between positive ones, to a net orientation of interfacial water at uncharged surfaces, with hydrogens pointing toward the bulk~\cite{wang_2024_charge, behjatian_2022_nonmonotonic, kubincova_2020_interfacial}, precisely the H-down offset resolved here. More generally, the offset provides an electrostatic baseline against which extrinsic contributions to charge asymmetry can be identified, from electrowetting, electrofreezing and the asymmetric differential capacitance of metal electrodes at macroscopic interfaces to the interactions of like-charged colloids and the stability and solubility of proteins.

\section{Conclusions}

Our combined experimental and theoretical analysis reveals how interfacial water reorganizes in response to surface charge at a weakly interacting, planar electrode. Direct spectral measurements of the absolute orientation of interfacial water show that water is not randomly oriented at zero net surface charge: it retains a net H-down alignment that arises from interfacial symmetry breaking and dispersive interactions rather than from electrostatics.
On top of this charge-independent offset, the field-induced reorientation
varies linearly and almost perfectly charge-symmetrically with surface charge
in both experiments and atomistic simulations.
This response holds over a wide range of practically relevant charge densities, independent of how the surface charge is introduced and whether it is localized or delocalized. 
Because of this offset, the potential of zero water orientation is displaced from the potential of zero charge: charging of one sign reinforces the pre-existing alignment, whereas charging of the other sign must first cancel it.
Absolute interfacial water orientation is therefore intrinsically charge-asymmetric, even without specific ion adsorption or surface chemistry. 
Charge asymmetry and a nonlinear response are thus distinct phenomena: here, the asymmetry stems from the zero-charge offset, while the response itself remains linear.
Separating this offset from the linear electrostatic response provides a reference against which extrinsic contributions, such as specific ion adsorption, surface chemistry, and surface heterogeneity can be identified. 
Together, experiments and simulations provide a unified microscopic picture of water organization at planar charged interfaces and a practical foundation for modeling aqueous interfaces.
These insights into interfacial water behavior have direct implications for aqueous systems spanning electrochemistry, catalysis, membrane transport, and geochemical processes.

\backmatter

\bmhead{Supplementary information}

This article has an accompanying Supplementary Information file containing Figures S1--S10.

\bmhead{Acknowledgments}

We are grateful for the financial support from the MaxWater Initiative of the Max Planck Society. Funded by the European Union (ERC, n-AQUA, 101071937). Views and opinions expressed are however those of the author(s) only and do not necessarily reflect those of the European Union or the European Research Council Executive Agency. Neither the European Union nor the granting authority can be held responsible for them. Y.W. acknowledges financial support from the Carl-Zeiss-Stiftung. S.J.C. is a Royal Society University Research Fellow at Durham University (URF\textbackslash R1\textbackslash 211144).

\section*{Declarations}

\bmhead{Funding}

MaxWater Initiative of the Max Planck Society; European Union (ERC, n-AQUA, 101071937); Carl-Zeiss-Stiftung (Y.W.); Royal Society University Research Fellowship, Durham University, URF\textbackslash R1\textbackslash 211144 (S.J.C.).

\bmhead{Competing interests}

The authors declare no competing interests.

\bmhead{Author contributions}

All authors designed the research; Y.L. performed the simulations, and Y.W. performed the experiments. Y.L. and Y.W. analyzed the data. All authors contributed to writing and revising the manuscript.

\section{Methods}

\bmhead{Molecular dynamics simulations}

We performed molecular dynamics simulations with the LAMMPS package~~\cite{thompson_lammps_2022}. The system consisted of a 1~M NaCl aqueous solution confined between two solid interfaces, each formed by a single graphene sheet and three CaF$_2$ layers. The setup followed Ref.~~\cite{wang_2026_wetting}, and the main parameters are summarized here.

Each graphene sheet contained 960 carbon atoms with a C–C bond length of 1.426~\AA. The simulation box (49.400~\AA~$\times$~51.336~\AA~$\times$~85.000~\AA) employed periodic boundary conditions in all directions, with 3000 water molecules confined between graphene layers separated by 40~\AA. 
Substrate and graphene atoms were kept fixed. 
Notably, our simulations do not include confined water between graphene and the substrate.
Localized surface charge densities of 0, $\pm$4.7, $\pm$9.5, $\pm$14.2, and $\pm$19.0~mC/m$^2$ were applied.
Unless stated otherwise, we modeled water with SPC/E~\cite{Berendsen_JCP_1987}, and we took ion parameters from the ECC-rescaled force fields of the Jungwirth group.~~\cite{Kohagen_JPCB_2016,Pluharova_MolPhys_2014} Lennard-Jones parameters for carbon followed Ref.~~\cite{Rafiee_NatMat_2012}.
Delocalized surface charging was modeled using the Siepmann–Sprik constant-potential method~~\cite{siepmann_influence_1995} implemented in the Electrode package~~\cite{ahrens-iwers_electrode_2022} with a reciprocal Gaussian width of 1.979~\AA$^{-1}$. 
Localized-charge simulations were performed at zero applied potential, allowing carbon charges to fluctuate.
For delocalized-charge simulations, potentials up to $\pm$0.5~V were applied in 0.05~V increments, corresponding to surface charge densities between 0 and $\pm$30~mC/m$^2$.

All simulations were run in the NVT ensemble at 300~K using a Nosé–Hoover thermostat (200~fs damping), a 1~fs timestep, and SHAKE constraints on O–H bonds. Initial configurations were generated with Packmol~~\cite{Martínez_JCC_2009}, equilibrated for 1~ns, and followed by 10~ns production runs for data collection.

\bmhead{Sample Preparation and Electrochemical HD-VSFG Measurements}

The CaF$_2$-supported graphene samples were prepared using a well-established polymer-assisted wet transfer method with commercial CVD-grown graphene, following previously reported procedures ~\cite{Montenegro_Nat_2021,Wang_Nature_2023,Wang_Angewandte_2023}. 
CaF$_2$ windows (25 mm diameter and 2 mm thickness, PI-KEM Ltd.) and graphene films (Grolltex Inc.) served as the optical substrate and working electrode, respectively.
The samples were integrated into a homemade PTFE electrochemical flow cell connected to a syringe pump, as described previously ~\cite{Wang_FaradayDiscuss_2024, Wang_Nature_2023,Wang_Angewandte_2023}. To minimize memory effects associated with interfacial restructuring ~\cite{wang_2024_aqueous, Wang_Nature_2023}, all measurements were conducted under continuous flow conditions with a linear flow velocity of approximately 1 mm s$^{-1}$.
The 2 M NaCl solutions were prepared by dissolving NaCl (Sigma-Aldrich) directly in deionized water supplied by a Milli-Q purification system 
(resistivity $\geq$ 18.2 M$\Omega$cm; TOC $\leq$ 4 ppb).
The solutions were allowed to equilibrate with atmospheric CO$_2$ for at least 1 h before measurement.
The electrochemical potential was controlled using a three-electrode configuration, with graphene, Au, and Pd/H$_2$ serving as the working, counter, and reference electrodes, respectively.

All the complex $\chi^{(2)}$ spectra were acquired using an HD-VSFG spectrometer in a non-collinear beam geometry, powered by a Ti:Sapphire regenerative amplifier laser system operating at a central wavelength of 800 nm, with a pulse width of 40 fs, pulse energy of 5 mJ, and a repetition rate of 1 kHz, as detailed in refs. ~\cite{Wang_Nature_2023, Seki_Chem_2021}.
In this setup, part of the laser output was sent to a grating-based pulse shaper with a cylindrical lens to generate a narrowband visible beam (10 cm$^{-1}$ FWHM), while the remaining output was directed to an optical parametric amplifier (TOPAS-C, Light Conversion) equipped with a GaSe crystal to produce a broadband infrared pulse. For heterodyne detection, the local oscillator (LO) signal was generated by focusing the IR and visible beams onto a 200 nm-thick ZnO film coated on a 1 mm-thick CaF$_2$ substrate, following established procedures~~\cite{extending_JPCC_2016}.
The LO, together with the IR and visible beams, was then directed and refocused using two pairs of off-axis parabolic mirrors to achieve spatial and temporal overlap at the CaF$_2$-supported graphene/water interface. The incident angles for the IR, visible, and LO beams (in CaF$_2$) were 33°, 39°, and 37.6°, respectively. HD-VSFG measurements were conducted at the ssp polarization combination, where 'ssp' refers to s-polarized SFG, s-polarized visible, and p-polarized IR beams. The measurements were performed in a dry air atmosphere to prevent spectral distortion from water vapor.

The phase measurement method followed previously established procedures ~\cite{Wang_Nature_2023,Wei_JACS_2023}. Briefly, a ~100 nm-thick gold layer was coated on half of the CaF$_2$-supported graphene sample and used as a 
reference. The phase of the gold-coated reference was determined by measuring the CaF$_2$-supported graphene/D$_2$O signal at the same position used for the CaF$_2$-supported graphene/water measurements, under the assumption that the response of the CaF$_2$-supported graphene/D$_2$O interface is purely real~~\cite{Yamaguchi_JCP_2011}. A height displacement sensor (CL-3000, Keyence) was used to check the sample height. 

The graphene charge density was determined from the potential-dependent shift of the Raman G-band frequency relative to its charge-neutral position (1,582~cm$^{-1}$), following previously established protocols~\cite{das_2008_monitoring, Wang_Nature_2023, Montenegro_Nat_2021}. Net surface charge densities were determined from the difference between the Im$\chi^{(2)}$ spectra measured at 10~mM and 2~M NaCl (Supplementary Fig.~S1), using a Gouy--Chapman--Stern model and following established protocols~\cite{wen_unveiling_2016, Wang_Nature_2023, Seki_Chem_2021}. In this analysis, we assumed a constant $\chi^{(3)}$ spectrum and a surface contribution independent of salt concentration, so that the difference spectrum isolates the diffuse-layer $\chi^{(3)}$ term. The horizontal error bars in Fig.~2a represent the standard deviation of the extracted charge across the O--H stretching region. 

\bmhead{Data availability}

All data required to reproduce the findings of this study will be made available in a public repository upon acceptance of this manuscript.

\bibliography{references,zotero2}

\end{document}


\title[Supplementary Information]{\textbf{Supplementary Information for}:\\ \\ Interfacial water responds linearly to charge yet is charge-asymmetric}

\author*[1,2]{\fnm{Yair} \sur{Litman}}\email{litmany@mpip-mainz.mpg.de}

\author*[1,2]{\fnm{Yongkang} \sur{Wang}}\email{wangy3@mpip-mainz.mpg.de}

\author[3]{\fnm{Stephen J.} \sur{Cox}}

\author*[1,2]{\fnm{Mischa} \sur{Bonn}}\email{bonn@mpip-mainz.mpg.de}

\affil[1]{\orgname{Max Planck Institute for Polymer Research}, \orgaddress{\street{Ackermannweg 10}, \city{Mainz}, \postcode{55128}, \country{Germany}}}

\affil[2]{\orgname{Max Planck Initiative for Liquid Research}, \orgaddress{\street{Ackermannweg 10}, \city{Mainz}, \postcode{55128}, \country{Germany}}}

\affil[3]{\orgdiv{Department of Chemistry}, \orgname{Durham University}, \orgaddress{\street{South Road}, \city{Durham}, \postcode{DH1 3LE}, \country{United Kingdom}}}

\maketitle
\renewcommand{\thefigure}{S\arabic{figure}}
\setcounter{figure}{0}

\section{Additional Figures}

\begin{figure}
\centering
\includegraphics[width=0.8\textwidth]{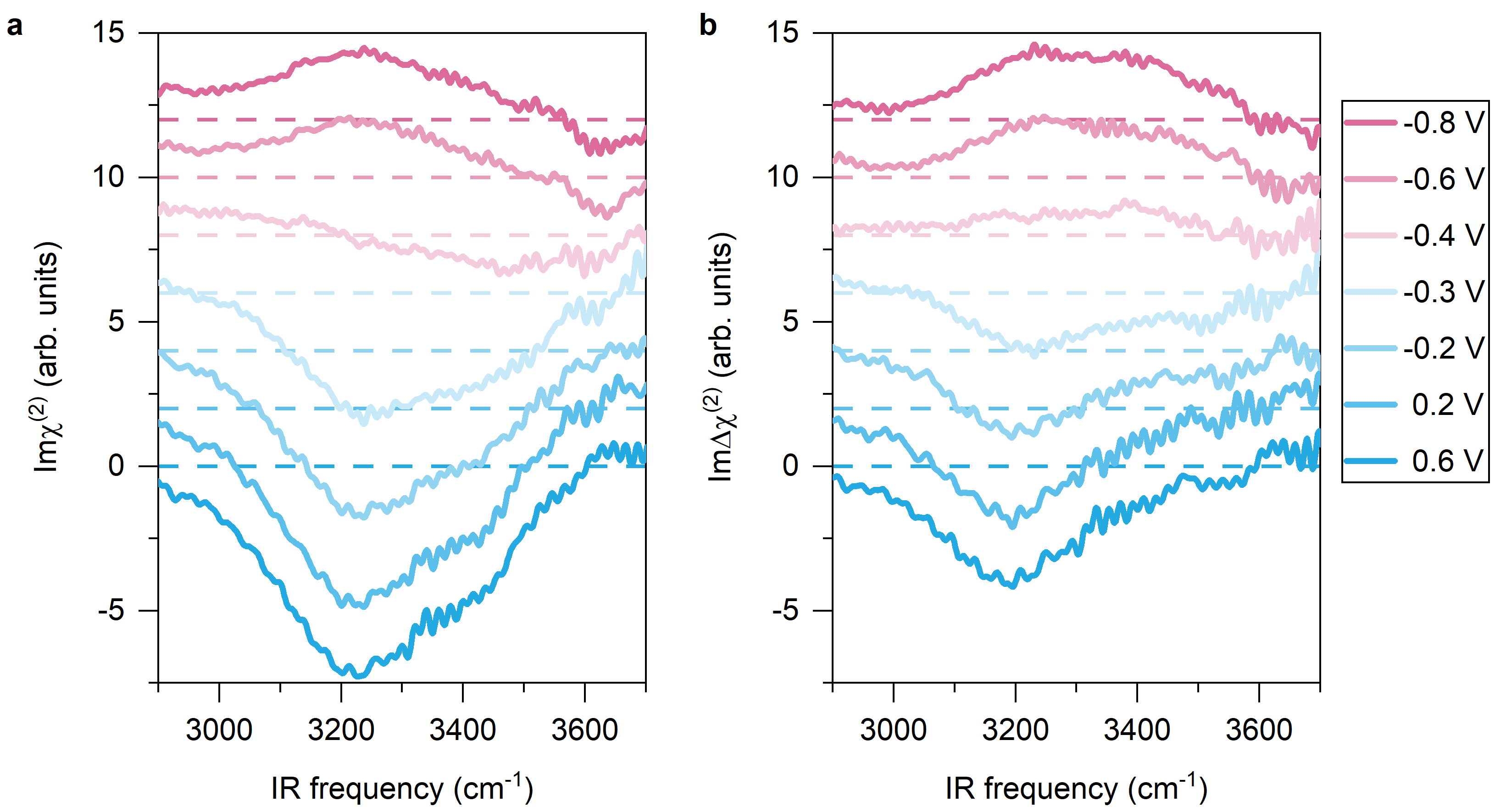}
  \caption{HD-SFG spectra of graphene/water interfaces at low ionic strength. (a) HD-SFG spectra of CaF$_2$-supported graphene/water interfaces measured in 10 mM NaCl at various applied potentials versus Pd/H$_2$. (b) Corresponding imaginary part of the differential second-order susceptibility, $\mathrm{Im}\Delta\chi^{(2)} = \mathrm{Im}[\chi^{(2)}$(10 mM) $-$ $\chi^{(2)}$(2 M)], used to determine the net surface charge density.
  }
  \label{fig:low_concentration}
\end{figure}

Figure~\ref{fig:low_concentration} shows HD-SFG spectra of CaF$_2$-supported graphene/water interfaces at various applied potentials versus Pd/H$_2$ in 10 mM NaCl, together with the corresponding imaginary part of the differential susceptibility, $\mathrm{Im}\Delta\chi^{(2)} = \mathrm{Im}[\chi^{(2)}$ (10 mM) $-$ $\chi^{(2)}$(2 M)], used to determine the net surface charge density.

\begin{figure}
\centering
\includegraphics[width=0.8\textwidth]{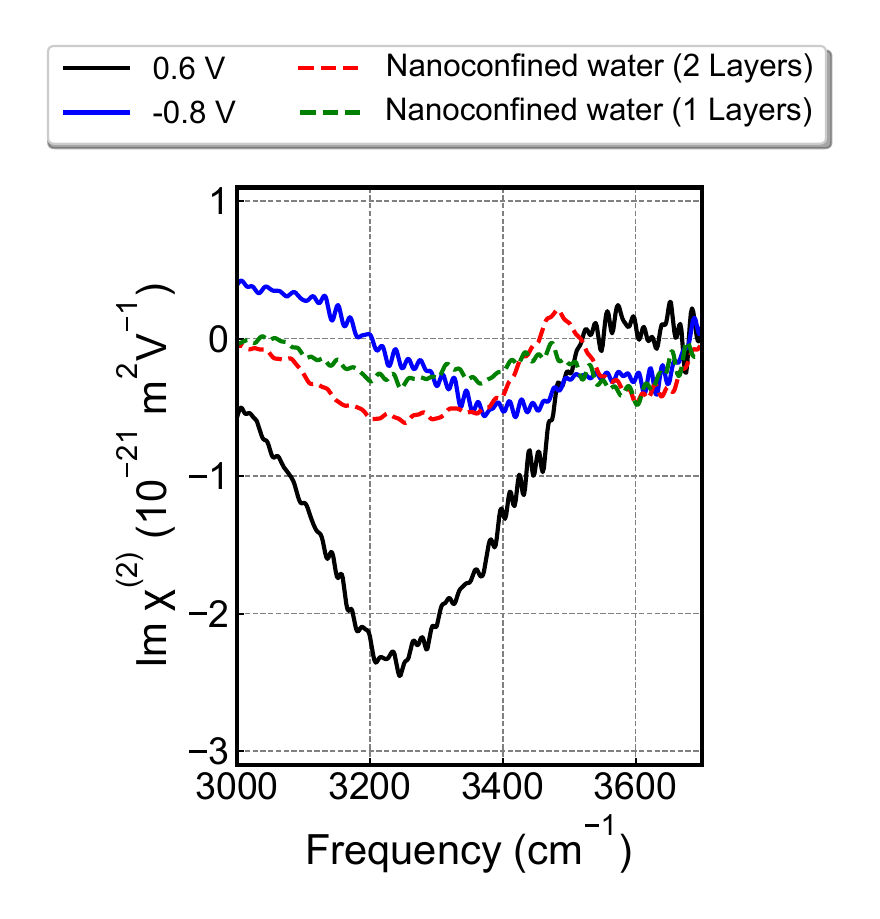}
  \caption{ Comparison of $\mathrm{Im}\,\chi^{(2)}$ spectra at selected voltage conditions (this work) with the signal obtained from one and two layers of water nanoconfined between CaF$_2$ and graphene (Ref.~\citenum{wang_NatComm_2025}). The two-layer spectrum sets the upper-limit uncertainty in Fig. 2a.}
  \label{fig:NC}
\end{figure}

Figure~\ref{fig:NC} compares the HD-SFG spectra of graphene/bulk solution interfaces with those of water nanoconfined between CaF$_2$ and monolayer graphene reported in Ref.~\citenum{wang_NatComm_2025}. As discussed in the main text, the contribution from nanoconfined water may be significant at large negative voltages (below –0.6 V), where the overall VSFG signal is small. At higher voltages, however—where the total signal increases—the relative contribution of the nanoconfined water becomes much less critical.

\begin{figure}
\centering
\includegraphics[width=.8\textwidth]{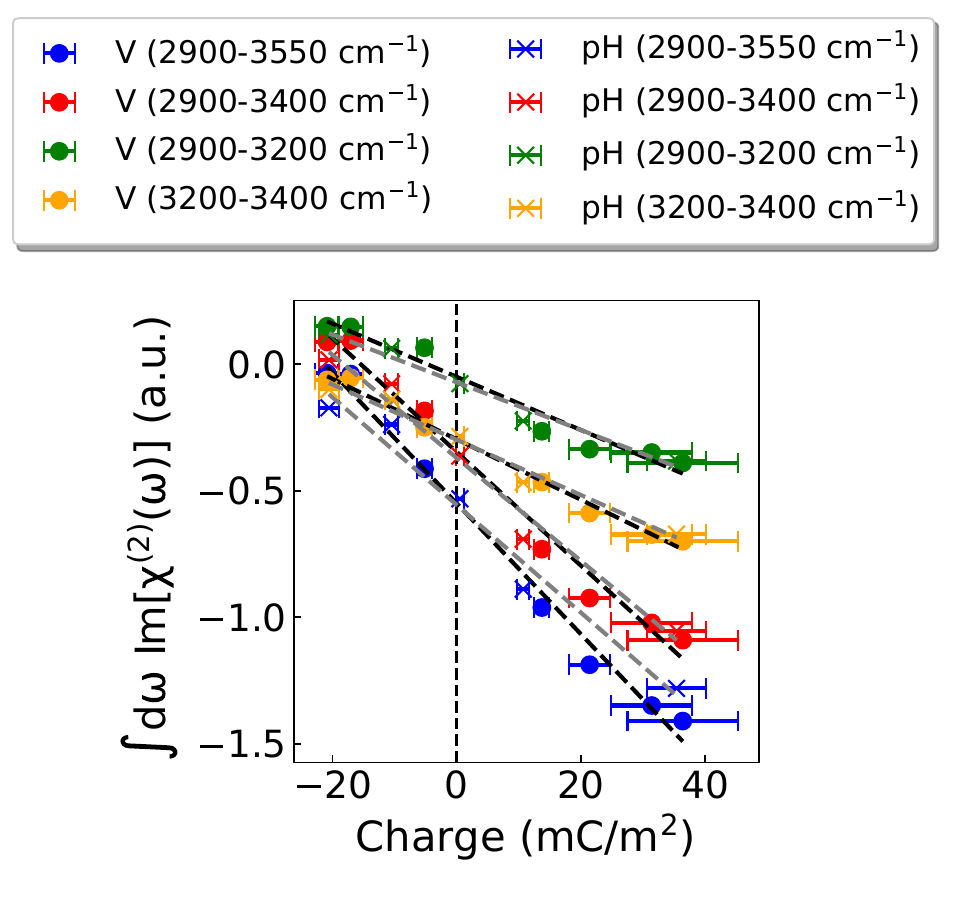}
  \caption{Comparison of the water response obtained by varying the frequency regions used to integrate the  Im $\chi^{(2)}$
 signal under different experimental conditions. Black and grey dashed lines represent linear fits for each integration region across measurements at different pH (crosses) and bias voltage (circles), respectively.}
  \label{fig:sfg-area}
\end{figure}

Figure~\ref{fig:sfg-area} shows the water orientation response as a function of surface charge, using different integration regions.
A linear trend is observed in all cases, showing the robustness of the conclusions discussed in the main text.

\begin{figure}
\centering
\includegraphics[width=0.8\textwidth,height= 8 cm,keepaspectratio]{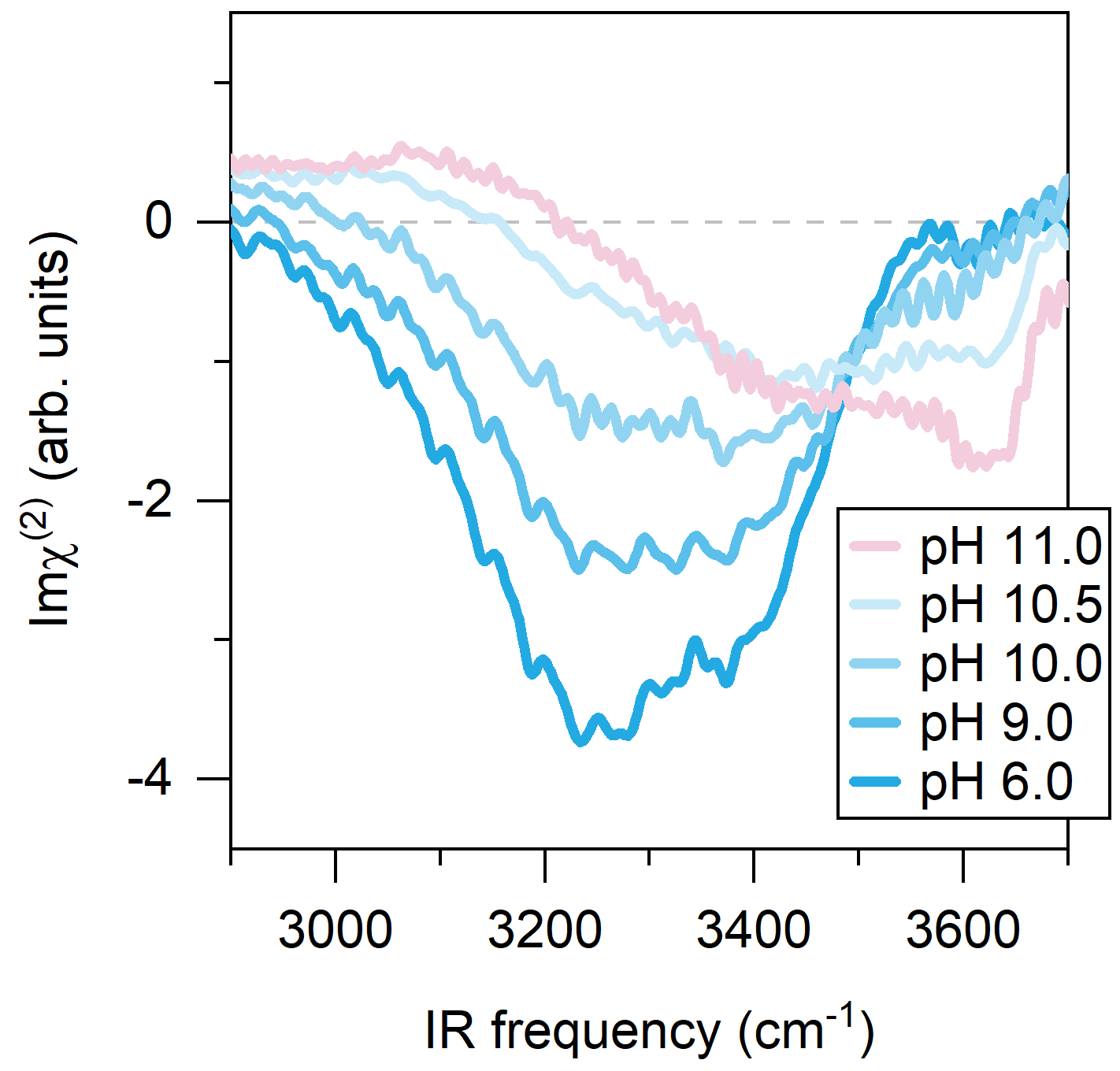}
  \caption{$\mathrm{Im}\chi^{(2)}$ spectra of the interfacial water O--H stretching region at CaF$_2$-supported graphene/water interfaces under various pH conditions.}
  \label{fig:ph_dependence}
\end{figure}

Figure~\ref{fig:ph_dependence} shows the $\mathrm{Im}\chi^{(2)}$ spectra of the interfacial water O--H stretching region at the CaF$_2$-supported graphene/water interface under various pH conditions. Similar spectral changes are observed upon varying the applied potential, as those shown in the main text.

\begin{figure}\centering
\includegraphics[width=0.8\textwidth,height= 8 cm,keepaspectratio]{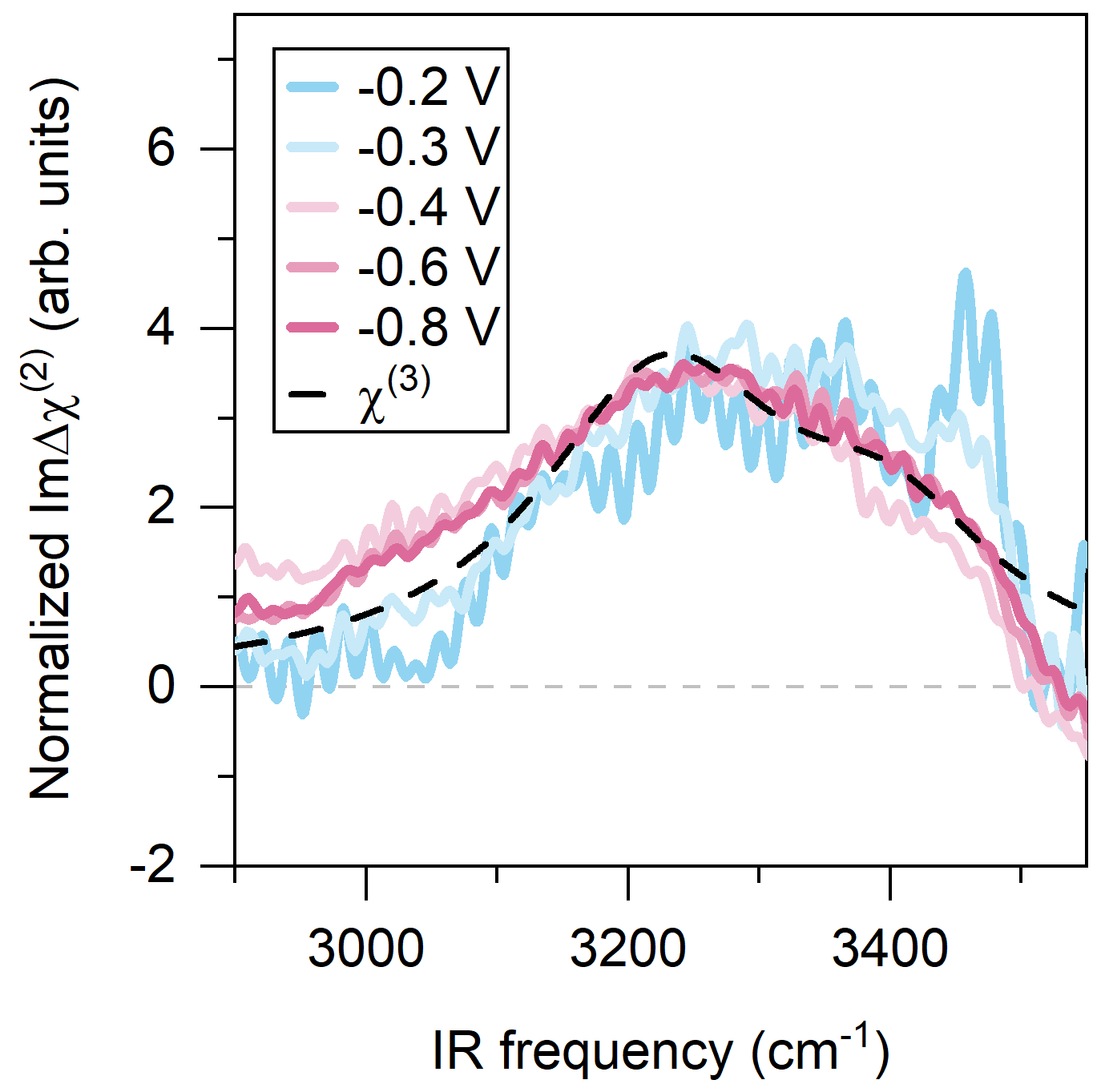}
  \caption{Normalized differential $\mathrm{Im}\Delta\chi^{(2)}$ spectra of interfacial water at the CaF$_2$-supported graphene/water interface under different applied potentials, with the spectrum at 0.6 V used as the reference. The spectrum at 0.2 V is not shown because the relative change is comparable to the experimental uncertainty. A rescaled $\chi^{(3)}$ spectrum is included for comparison. The grey dashed lines indicate the zero baselines.}
  \label{fig:normalized_differential_spectral}
\end{figure}

Figure~\ref{fig:normalized_differential_spectral} shows the normalized differential $\mathrm{Im}\Delta\chi^{(2)}$ spectra of interfacial water at the CaF$_2$-supported graphene/water interface under different applied potentials. After normalization, the spectra collapse onto a common lineshape that closely matches the field-induced $\chi^{(3)}$ response of bulk water reported previously\cite{wen_2016_unveiling, Wang_Nature_2023}.

\begin{figure}\centering
\includegraphics[width=1\textwidth]{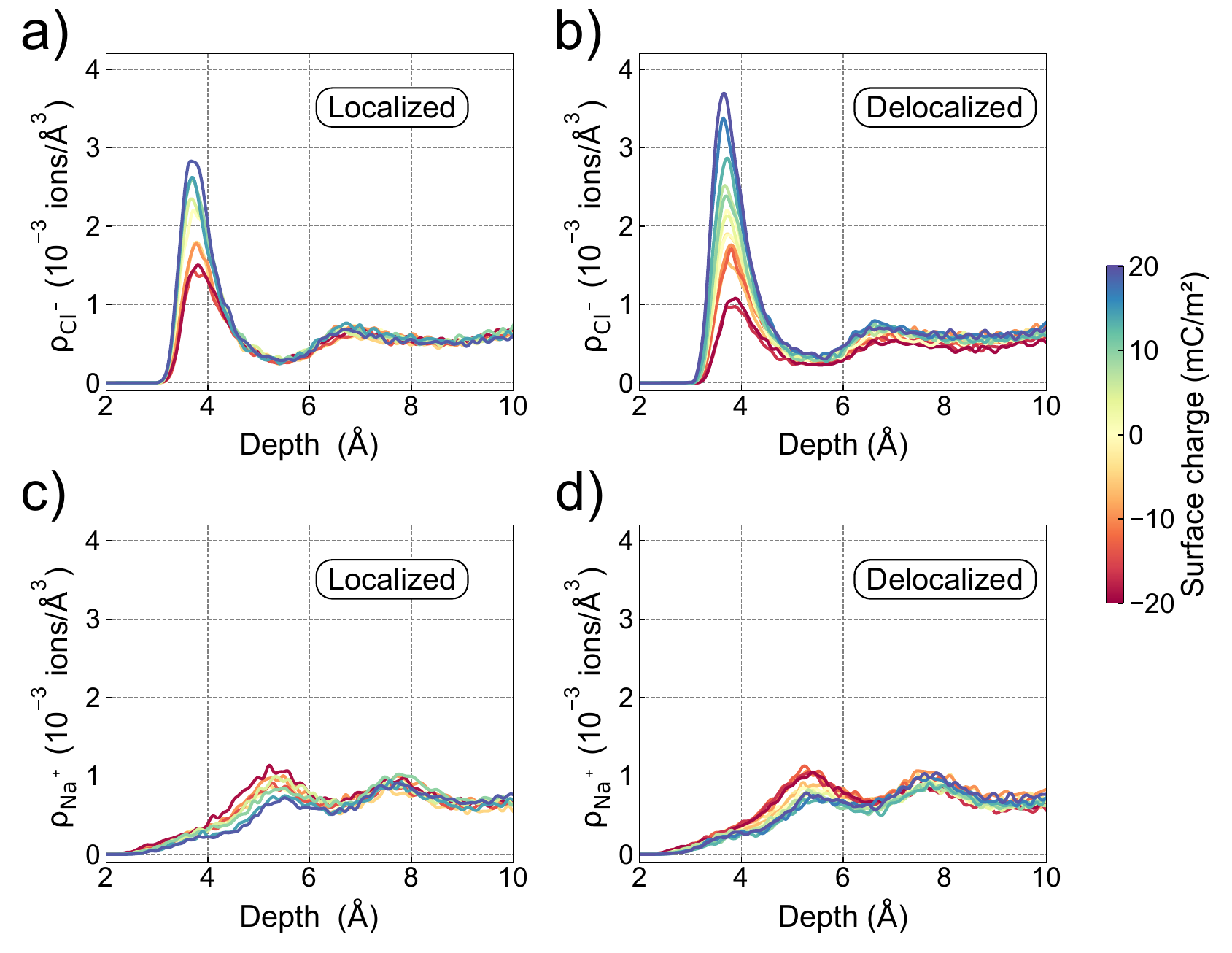}
  \caption{ Density profiles, $\rho$, at different surface charge densities for Na${^+}$ and  Cl${^-}$, respectively for localized and delocalized surface charges. 
In all cases, the zero depth is defined as the center of the carbon atoms.}
  \label{fig:ions-profile}
\end{figure}

Figure~\ref{fig:ions-profile} shows the Na${^+}$ and Cl${^-}$ ion density profiles for both localized and delocalized surface charge scenarios.
While the cation (Na${^+}$) profiles remain largely unaffected by the nature of the surface charges, the anion (Cl${^-}$) profiles exhibit approximately 20\% higher sensitivity in the case of delocalized charges.

\begin{figure}\centering
\includegraphics[width=0.8\textwidth,height=10cm,keepaspectratio]{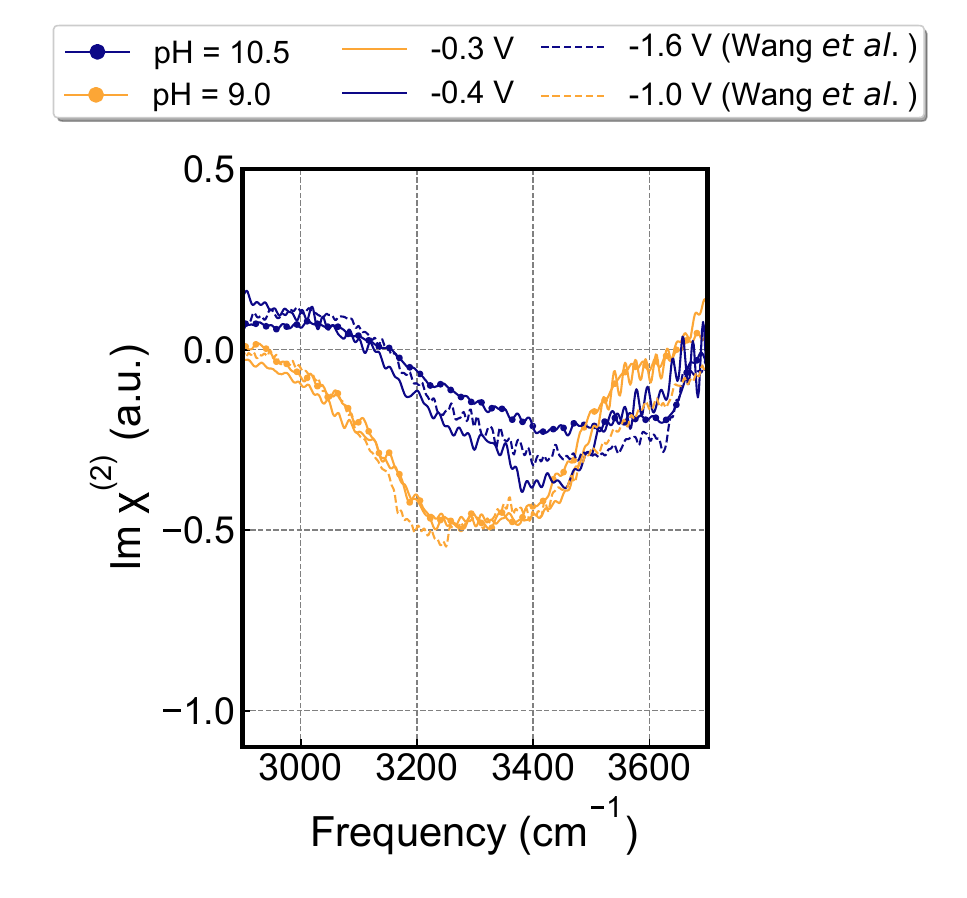}
  \caption{Comparison of $\mathrm{Im}\,\chi^{(2)}$ spectra at selected pH and voltage conditions between a 2 M NaCl aqueous solution (this work) and the inferred response of pure water from Ref.~\citenum{Wang_Nature_2023}. The pH and bias voltage values were selected to achieve comparable surface charge densities, estimated to be approximately $-10$~mC/m$^2$ and $10$~mC/m$^2$ for pH = 10.5 and pH = 9, respectively. The net surface charge densities are comparable between the two studies: $14$~mC/m$^2$ at $-0.3$~V in this work and $8$~mC/m$^2$ at $-1.0$~V in Ref.~\citenum{Wang_Nature_2023}, as well as $-5$~mC/m$^2$ at $-0.4$~V in this work and $4$~mC/m$^2$ at $-1.6$~V in Ref.~\citenum{Wang_Nature_2023}. The larger applied voltages required in Ref.~\citenum{Wang_Nature_2023} likely arise from the use of fast-flow conditions, which reduce the effective surface charging. A slight enhancement of the high-frequency response around 3400~cm$^{-1}$ is observed for the 2 M NaCl solution investigated here, which may originate from ion-specific effects; however, such contributions, if present, remain minor compared with the overall electrostatic response.}
  \label{fig:sfg-ions}
\end{figure}

Figure~\ref{fig:sfg-ions} compares the Im$\chi^{(2)}$ spectra presented in the main text, obtained using a 2 M electrolyte concentration, with the spectrum from Ref.~\cite{Wang_Nature_2023}, obtained for water without added electrolyte.
The similarity between the two spectra indicates the absence of specific ion effects.

\begin{figure}
    \centering
\includegraphics[width=0.8\textwidth,keepaspectratio]{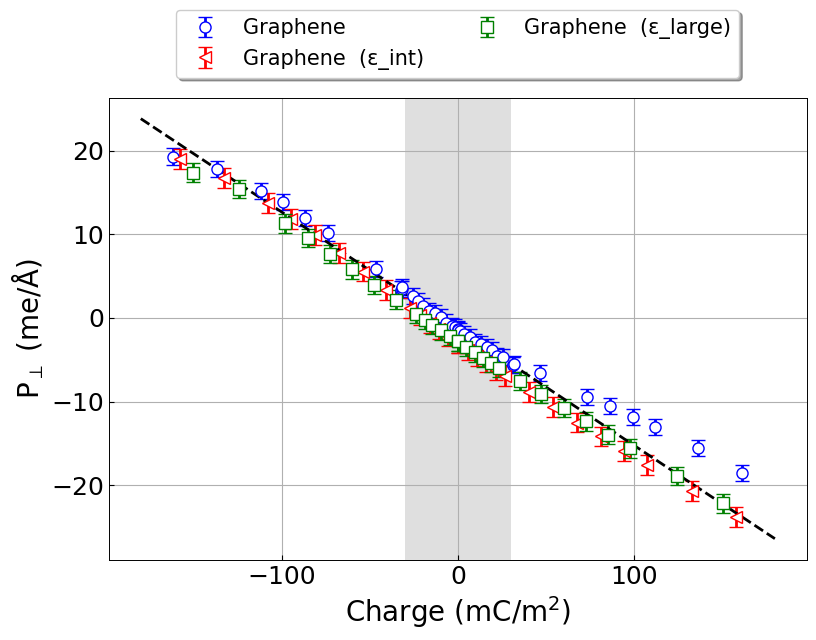}
    \caption{Orientation of interfacial water as a function of surface charge
    density $\sigma$, quantified by the density-weighted orientational order
    parameter
    $P_\perp = -\mu\int \rho(z)\langle  \cos\theta(z) \rangle\,dz$, where
    $\theta$ is the angle between the water dipole and the surface normal.
    Results are shown for CaF$_2$-supported graphene electrodes with three
    carbon-oxygen van der Waals well depths at fixed
    $\sigma_\mathrm{CO} = 3.283$~\AA:
    $\varepsilon_\mathrm{CO} = 0.115$~kcal/mol (blue circles, standard
    graphene), $\varepsilon_\mathrm{int} = 0.281$~kcal/mol (red triangles), and
    $\varepsilon_\mathrm{large} = 0.841$~kcal/mol (green squares). The highest value is representative of Ag(111) electrode\cite{Heinz2008}. The dashed line is the
    linear-response prediction obtained by fitting the data within the shaded
    region ($|\sigma| \lesssim 30$~mC/m$^2$); the shaded region indicates the
    range of surface charge densities over which the fit was performed. All three electrodes coincide within the
    linear-response regime and along the negative-charge branch; at positive
    charge densities above $\approx 60$~mC/m$^2$ the standard graphene
    electrode departs from linearity, while the two strongly interacting
    electrodes remain linear over the full range sampled.}
    \label{fig:high_charge_density}
  \end{figure}

  To assess how robust this linearity is to the details of the water-electrode
  interaction, we artificially increased the carbon-oxygen van der Waals well
  depth from its graphene value to one representative of a silver electrode, together with an intermediate value (see caption for
  details). Figure~\ref{fig:high_charge_density} shows that all three model electrodes
  respond linearly and are indistinguishable within the shaded region ($\pm 30$ mC/m$^2$). At higher
  surface charge densities, the bare graphene electrode departs from linearity,
  whereas the more hydrophilic electrodes remain linear over a substantially
  wider range of surface charge densities.

\begin{figure}
    \centering
    \includegraphics[width=0.8\textwidth,keepaspectratio]{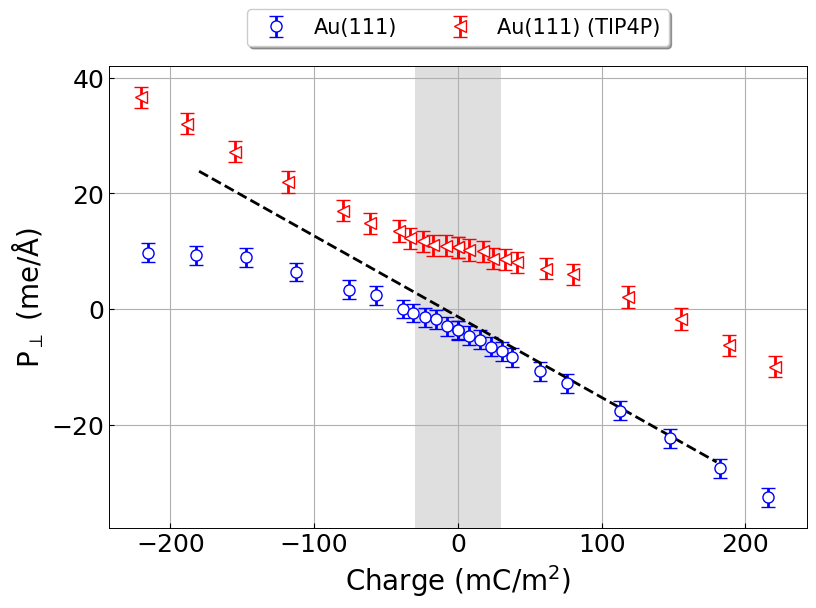}
    \caption{
Orientation of interfacial water,
    $P_\perp$ as defined in Fig. \ref{fig:high_charge_density},
    as a function of surface charge
    density, $\sigma$, at the Au(111) surface.  Blue circles:
    SPC/E water; red triangles: TIP4P/2005 water. Both data sets use the same
    Au(111) electrode (12--6 Lennard-Jones parameters of Ref.~\cite{Heinz2008}),
    so that they differ only in the water model. The dashed line
    is the linear-response prediction obtained from the SPC/E data within the
    shaded region ($|\sigma| \lesssim 30$~mC/m$^2$). Both
    water models are linear within the shaded region and depart from linearity
    beyond it, but the character of the departure differs between them and from
    that of the graphene electrodes of Fig.~\ref{fig:high_charge_density}.}
    \label{fig:metal}
  \end{figure}

  We next verified that these observations are not specific to graphitic electrodes. Figure~\ref{fig:metal} shows the same analysis for Au(111): the   response remains linear within the shaded region, but the way in which linearity
  breaks down beyond it differs markedly from graphene, the SPC/E interface saturating under negative charging while remaining linear under positive charging. Repeating the calculation with the TIP4P/2005 water model yields the same qualitative behavior, namely a linear regime at low surface charge densities followed by a departure from linearity at higher ones, although the
  slope and the value of $P_\perp$ at the point of zero charge differ. Linearity
  at low surface charge densities is therefore robust with respect to both the
  electrode material and the water model, whereas the onset and the character of the nonlinear response at high surface charge densities are not.

\begin{figure}\centering
\includegraphics[width=0.8\textwidth,height= 8 cm,keepaspectratio]{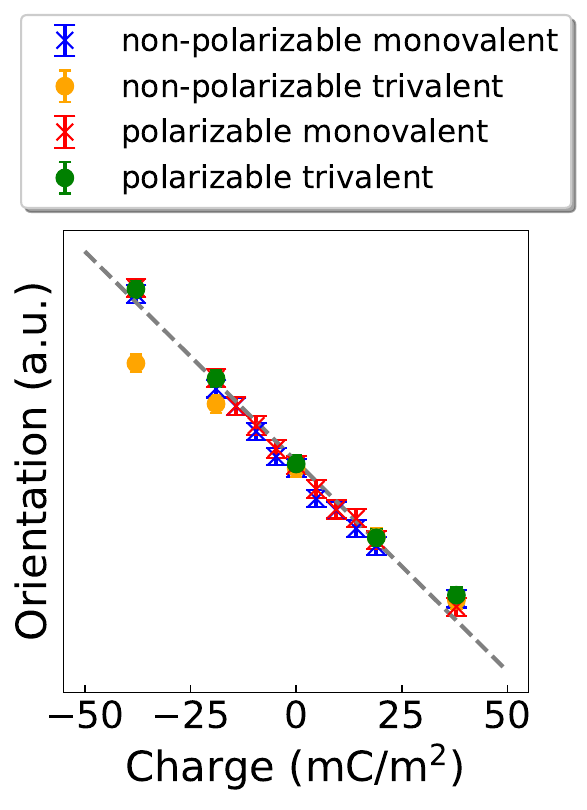}
  \caption{Influence of ion multivalency and graphene polarization on the water average response.  The water response is linear over $\pm$25 mC/m$^2$, with negligible dependence on graphene polarizability or ion valence, highlighting the robustness of interfacial water behavior. The dashed line serves as a guide to the eye. Nonlinear effects emerge at higher (negative) charge densities for non-polarizable graphene trivalent ions.}
  \label{fig:multivalent}
\end{figure}

The simulations presented in the main text have employed localized monovalent charges exclusively. To test the robustness of our findings, we performed additional simulations using trivalent localized charges on the substrate. Figure~\ref{fig:multivalent} shows the water orientational response for monovalent and trivalent surface ions. A deviation from linearity is observed for negatively charged localized trivalent ions below -25 mC/m$^2$. As a further test of the role of graphene polarizability, we performed simulations using a non-polarizable graphene model. These results indicate that graphene’s polarizability does not play a significant role in observed linearity for localized charge densities smaller than $\pm$ 25 mC/m$^2$. Still, it reduces the nonlinear behavior in the presence of negatively charged trivalent ions. As highly charged, unscreened trivalent ions are prone to hydrolysis at realistic aqueous interfaces, these computational results further substantiate the robustness of the linear water response at small and moderate surface charge densities.

\FloatBarrier
\pagebreak


\FloatBarrier
\pagebreak

\bibliography{references,zotero2}